\documentclass[twocolumn,showpacs,amsmath,amssymb, showkeys]{revtex4}
\usepackage{graphicx}
\usepackage[tight]{subfigure}
\usepackage{dcolumn}
\usepackage{bm}

\begin{document}

\title{Comment on \textquotedblleft Impact of a Global Quadratic
Potential on Galactic Rotation Curves\textquotedblright }

\author {Kamal K. Nandi$^{1,2,a}$ and Arunava Bhadra$^{3,b}$}

\affiliation{$^{1}$ Ya.B. Zel'dovich International Center for Astrophysics, Ufa 450000, Russia \\
$^{2}$ Department of Mathematics, University of North Bengal, Siliguri 734013, India \\
$^{3}$ High Energy $\&$ Cosmic Ray Research Centre, University of North Bengal, Siliguri 734013, India \\
$^{a}$ kamalnandi1952@yahoo.co.in , \\
$^{b}$ aru\_bhadra@yahoo.com \\ }

\begin{abstract}
Conformal gravity theory can explain observed flat rotation curves of galaxies without invoking hypothetical dark matter. Within this theory, we obtain a generic formula for the sizes of galaxies exploiting the stability criterion of circular orbits. It is found that different galaxies have different finite sizes uniquely caused by the assumed quadratic potential of cosmological origin. Observations on where circular orbits might actually terminate could thus be very instructive in relation to the galactic sizes predicted here. 
  
\end{abstract}
\maketitle

In a recent Letter, Mannheim and O'Brien [1] have presented conformal
gravity rotational velocity\ $v_{\text{total}}^{2}$ fits to the observed
data of several galaxy samples, which seem good enough to indicate that
conformal gravity could be an interesting alternative to dark matter
hypothesis. An important prediction of the theory is the testable upper
limit on the size of the galaxies projected from $v_{\text{total}%
}^{2}\rightarrow 0$ (hence effectively the global limit $R_{\text{proj}}^{%
\text{global}}\leq \gamma _{0}/\kappa \approx 100$ Kpc). The purpose of this
Comment is to correct that this upper limit should be fixed by the criterion
of stability of orbits. If the canonical stable limit is observationally
surpassed, conformal theory would be falsified even if the last observed
orbit remains within $R_{\text{proj}}^{\text{global}}$.

Note that emission occurs from stable circular material orbits with
information propagating along null geodesics (see, for instance [2]). The
stability criterion can severely constrain the extent of the H1 gas and we
observe that conformal gravity endows each galaxy with a maximal stable
limit $R=R_{\text{stable}}^{\max }$ that falls within the limit $R_{\text{%
proj}}^{\text{global}}$. The two limits often differ significantly, by as
much as 20-30\%. For instance, with $UGC2885$, we find $R_{\text{stable}%
}^{\max }=191$ Kpc but $R_{\text{proj}}^{\text{global}}=253$ Kpc and this
difference ($\sim 62$ Kpc) could be well distinguished. Since stability is
an essential physical condition, we think that only $R_{\text{stable}}^{\max
}$ should be regarded as the testable upper limit on the size of a galaxy.
With their metric ansatz, the geodesic for a single test particle yields the
tangential velocity for circular orbits $v^{2}=\left( Rc^{2}/2\right)
B^{\prime }$ (primes denote derivatives with respect to $R$). With
approximate $v_{\text{total}}^{2}$, it integrates to 

\begin{eqnarray}
B(R) &=&1-\frac{2N^{\ast }\beta ^{\ast }}{R}+\left( N^{\ast }\gamma ^{\ast
}+\gamma _{0}\right) R-\kappa R^{2}+  \nonumber \\
&&\frac{3R_{0}^{2}N^{\ast }\gamma ^{\ast }}{2R}+\frac{15R_{0}^{4}N^{\ast
}\gamma ^{\ast }-24R_{0}^{2}N^{\ast }\beta ^{\ast }}{8R^{3}}.
\end{eqnarray}%

The radial geodesic is given by%

\begin{equation}
\left( \frac{dR}{dt}\right) ^{2}=B^{2}(R)-a\frac{B^{3}(R)}{R^{2}}-bB^{3}(R),
\end{equation}%

where $a$ \ and $b$ are constants fixed by the usual conditions for circular
orbits. The condition for stability is that the second derivative of the
right hand side (\textquotedblleft effective potential\textquotedblright )
of Eq.(2) with respect to $R$ must be negative, which leads to the generic
requirement that%

\begin{equation}
f(R)\equiv 2B^{\prime 2}(R)-B(R)B^{\prime \prime }(R)-3B(R)B^{\prime
}(R)/R<0.
\end{equation}

We illustrate our comments here only for $UGC2885$ (Figs.1a,b). Rest of the
samples yield similar patterns. The very fact that there exists a finite
limit $R_{\text{stable}}^{\max }$, caused \textit{entirely }by the quadratic
potential $V_{\kappa }(r)=-\kappa c^{2}r^{2}/2$, clearly distinguishes
conformal theory from some dark matter models because in the latter there is
no such limit, see e.g., [3,4]. Note that we don't know precisely what would happen beyond this special radius 
$R_{\text{stable}}^{\max }$, but gas in non-circular motions at larger radii is not certainly excluded. Interestingly, the predicted $R_{\text{stable%
}}^{\max }$ does not even much exceed the current $R_{\text{last}}$ for many
samples, e.g., $UGC0128$ has $R_{\text{stable}}^{\max }=65.6$ Kpc, while $R_{%
\text{last}}=54.8$ Kpc, so we might not have to wait too long. The main
thing to watch is whether or not any updated $R_{\text{last}}$ shoots past $%
R_{\text{stable}}^{\max }$, which fortunately has not happened yet. Updated
observations on $R_{\text{last}}$ would thus provide a nice test of
conformal gravity prediction of $R_{\text{stable}}^{\max }$ and hence of the
global quadratic potential.

\begin{figure}[ht]
\centering
\subfigure[] 
{\includegraphics[scale=0.7]{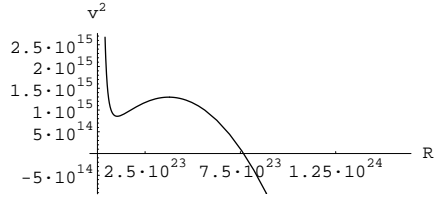}} 
\subfigure[] 
{\includegraphics[scale=0.7]{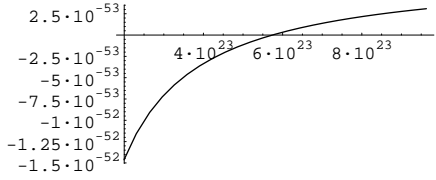} }
\caption{ (a) $v_{\text{total}}^{2}$ vs $R$ and (b) $f(R)$ vs $R$ for
UGC2885 }
\end{figure}

\noindent Dr. Mannheim [5] has the opinion that the general stability analysis may be performed considering many body dynamics. 
Observations on where circular orbits might actually terminate could thus be very instructive. \newline


\begin{thebibliography}{99}

\bibitem{ab:1} P.D. Mannheim and J.G. O'Brien, Phys. Rev. Lett. 
\textbf{106}, 121101 (2011).\newline
\bibitem{ab:2} K. Lake, Phys. Rev. Lett. \textbf{92}, 051101 (2004).%
\newline
\bibitem{ab:3} K. K. Nandi\textit{\ et al}, Mon. Not. Roy.
Astron. Soc. \textbf{399}, 2079 (2009). \newline
\bibitem{ab:4} F. Rahaman \textit{et al}, Phys. Lett. B \textbf{694},
10 (2010). \newline
\bibitem{ab:5} Private correspondence
\end{thebibliography}
\end{document}